\input harvmac
\input epsf
%\sequentialequations

\def\deight{\delta^8(x-x_0)}
%%%%%%%%%%%%%%%%%%%%%%%%%%references%%%%%%%%%%%%%%%%%%%%%%%%%%%%
\lref\giveon{A. Giveon and D. Kutasov, ``Brane Dynamics and Gauge
Theory,''
hep-th/9802067.}
\lref\karch{A. Karch, ``Field Theory Dynamics from Branes in String
Theory,''
hep-th/9812072.}
\lref\hanany{A. Hanany and E. Witten, ``Type IIB Superstrings, BPS
Monopoles,
and Three
Dimensional Gauge Dynamics,'' Nucl. Phys. B492 (1997) 152,
hep-th/9611230.}
\lref\witten{E. Witten, ``Solutions of Four-Dimensional Field Theories
Via
M-Theory,''
Nucl. Phys. B500 (1997) 3, hep-th/9703166.}
\lref\seiberg{N. Seiberg and E. Witten,
``Electric-Magnetic Duality, Monopole Condensation and Confinement
in $N=2$ Supersymmetric Yang-Mills Theory,'' Nucl. Phys. B426 (1994) 19,
Erratum ibid. B430 (1994) 485,
hep-th/9407087; ``Monopoles, Duality and Chiral Symmetry Breaking in
$N=2$
Supersymmetric QCD,''
Nucl. Phys. B431 (1994) 484, hep-th/9408099.}
\lref\aaz{R. Khuri,
Phys. Rev. D48(1993) 2947.}
\lref\baz{G. Papadopoulos, P.K. Townsend,
 Phys. Lett. B380 (1996) 273, hep-th/9603087.}
\lref\caz{ A.A. Tseytlin,
 Nucl.Phys. B475 (1996) 149, hep-th/9604035.}
\lref\daz{I.R. Klebanov, A.A. Tseytlin,
 Nucl. Phys.  B475 (1996) 179, hep-th/9604166.}
\lref\eaz{K. Behrndt, E. Bergshoeff, and B. Janssen,
 Phys. Rev.  D55 (1997) 3785, hep-th/9604168.}
\lref\faz{J. P. Gauntlett, D. A. Kastor, and J. Traschen,
Nucl. Phys. B478 (1996) 544, hep-th/9604179.}
\lref\gaz{E. Bergshoeff, M.
de Roo, E. Eyras, B. Janssen, and J. P. van der Schaar,
Nucl. Phys.  B494 (1997) 119, hep-th/9612095.}
\lref\haz{A.A. Tseytlin,
 Mod. Phys. Lett. A11 (1996) 689, hep-th/9601177.}
\lref\iaz{A.A. Tseytlin,
 Class. Quant. Grav. 14 (1997) 2085, hep-th/9702163.}
\lref\jaz{K. Behrndt and M. Cvetic,   Phys. Rev.
 D56 (1997) 1188, hep-th/9702205.}
\lref\kaz{J.C. Breckenridge, G. Michaud, and R. C. Myers,
 Phys. Rev. D56 (1997) 5172, hep-th/9703041.}
\lref\laz{E. Bergshoeff, M. de Roo, E. Eyras, B. Janssen, J.P. van der
Schaar,
 Class. Quant. Grav. 14 (1997) 2757, hep-th/9704120.}
\lref\maz{V. Balasubramanian, F. Larsen, and R. G. Leigh,
 Phys. Rev. D57 (1998) 3509, hep-th/9704143.}
\lref\naz{G. Michaud and R.C. Myers,  Phys. Rev.
 D56 (1997) 3698, hep-th/9705079.}
\lref\nazz{R. Argurio, F. Englert and  L. Houart,  Phys. Lett. B398 (1997)
61, hep-th/9701042. }
\lref\nazw{I.Ya.Aref'eva, O.A.Rytchkov, hep-th/9612236.}
\lref\oaz{J. D. Edelstein, L. Tataru, and R. Tatar,
 JHEP 9806 (1998) 003, hep-th/9801049.}
\lref\paz{I. Ya. Aref'eva, M. G. Ivanov, O. A. Rytchkov, and
I. V. Volovich,  Class. Quant. Grav. 15 (1998) 2923,
hep-th/9802163.}
\lref\qaz{J. P. Gauntlett, R. C. Myers, and P. K.
Townsend, hep-th/9809065.}
\lref\yang{H. Yang, ``Localized Intersection Brane Solutions of $D=11$
Supergravity,''
hep-th/9902128.}
\lref\fayya{A. Fayyazuddin and D.J. Smith, ``Localized Intersections of
M5-Branes and Four-Dimensional Superconformal Field Theories,''
hep-th/9902210.}
\lref\raz{A. Loewy, hep-th/9903038.}
\lref\gauntlett{J.P. Gauntlett, ``Intersecting Branes,''
Lectures given at APCTP Winter School on Dualities of Gauge and String
Theories,
17-28 Feb 1997,
Seoul and Sokcho, Korea, hep-th/9705011.}
\lref\stelle{K.S. Stelle, ``BPS Branes in Supergravity,''
Talk given at ICTP Summer School in High-energy Physics and Cosmology,
Trieste,
Italy, 10 Jun -
26 Jul 1996 and at the ICTP Summer School in High-Energy Physics and
Cosmology Trieste, Italy, 2 Jun - 11 Jul 1997.
In *Trieste 1997, High energy physics and cosmology* 29-127,
hep-th/9803116.}
\lref\youmbh{D.~Youm,
``Black holes and solitons in string theory,"
hep-th/9710046.}
\lref\don{S. Surya and D. Marolf, ``Localized Branes and Black Holes,''
Phys. Rev. D58 (1998) 124013, hep-th/9805121.}
\lref\amdon{D. Marolf and A. Peet, ``Brane Baldness vs. Superselection
Sectors,'' hep-th/9903213.}
\lref\malda{J. Maldacena, ``The Large N Limit of Superconformal Field
Theories
and Supergravity,''
Adv. Theor. Math. Phys. 2 (1998) 231, hep-th/9711200.}
\lref\itzakhi{N. Itzakhi, J. Maldacena, J. Sonnenschein and S.
Yankielowicz,
``Supergravity
and the Large N Limit of Theories with Sixteen Supercharges,'' Phys.
Rev. D58
(1998) 046004,
hep-th/9802042.}
\lref\LP{H. L\"u and C.N. Pope, ``Interacting Intersections,''
hep-th/9710155.}
\lref\yang{H. Yang, ``Localized Intersection Brane Solutions of $D=11$
Supergravity,''
hep-th/9902128.}
\lref\youm{D. Youm, ``Localized Intersecting BPS Branes,''
hep-th/9902208.}
\lref\fayya{A. Fayyazuddin and D.J. Smith, ``Localized Intersections of
M5-Branes and
Four-Dimensional Superconformal Field Theories,'' hep-th/9902210.}
\lref\callan{C.G. Callan and J.M. Maldacena, ``Brane Dynamics
from the Born-Infeld Action,''
Nucl. Phys. B513 (1998) 198, hep-th/9708147.}
\lref\LT{L. Thorlacius, ``Born-Infeld String as a Boundary Conformal
Field
Theory,''
Phys. Rev. Lett. 80 (1998) 1588-1590,
hep-th/9710181}
\lref\gibbons{G.W. Gibbons, ``Born-Infeld Particles and Dirichlet
P-Branes,''
Nucl. Phys. B514 (1998) 603, hep-th/9709027.}
\lref\howe{P.S. Howe, N.D. Lambert and P.C. West, ``The Self-Dual String
Soliton,''
Nucl. Phys. B515 (1998) 203, hep-th/9709014.}
\lref\li{M. Li, ``'T Hooft Vortices on D-Branes,''
 JHEP 9807 (1998) 003, hep-th/9803252.}
\lref\geroch{R. Geroch and J. Traschen,
``Strings and Other Distributional Sources in General Relativity,''
Phys. Rev. D36 (1987) 1017.}
\lref\gr{I.S.  Gradsteyn and I. M. Ryzhik, ``Tables of Integrals,
Series, and Products'' (Academic Press, NY, 1965) (2.263).}
\lref\rub{P. Ruback,
{\it Commun. Math. Phys.} {\bf 116} (1988) 645.}
\lref\BT{ E. Bergshoeff and P.K. Townsend, hep-th/9904020.}
\lref\wald{R.M. Wald, ``General Relativity,'' University of Chicago
Press, 1984.}
\lref\ITY{ N. Itzhaki, A. Tseytlin, and S. Yankielowicz,
{\it Phys. Lett.}  { B432} (1998) 298, hep-th/9803103.  }
\lref\Juan{
J. Maldacena,
{\it Adv. Theor. Math. Phys.}  { 2} (1998) 231, hep-th/9711200.}
\lref\IMSY{
N. Itzhaki, J. M. Maldacena, J. Sonnenschein, and S.  Yankielowicz,
Phys. Rev. D58 (1998) 046004, hep-th/9802042.}
\lref\schmid{
C. Schmidhuber,
{\it Nucl. Phys.} { B467}, 146 (1996), hep-th/9601003.}
\lref\dt{
E. Bergshoeff and M. De Roo,
{\it Phys. Lett.} {B380}, 265 (1996), hep-th/9603123.}
%
%%%%%%%%%%%%%%%%%%%%%%%%%%%%%%%%%%%%%%%%%%%%%%%%%%%%%%%%%%%%%%

%%%%%%%%%%%%%%%%%%%%%%%%%%% Title Page %%%%%%%%%%%%%%%%%%%%%%%%%%%%%%%

\Title{\vbox{\baselineskip12pt
\hbox{NSF-ITP-99-30, UMHEP-458}
\hbox{SU-GP-99/5-1, hep-th/9905094}
}}
{\vbox{\centerline{\titlerm Fully Localized Brane Intersections}
\bigskip\centerline{\titlerm - The Plot Thickens}
 }}
{\baselineskip=12pt
\centerline{Andr\'es Gomberoff${}^{a,c,d}$, David Kastor${}^{a,b}$,
Donald
Marolf${}^{a,c}$ and Jennie
Traschen${}^{a,b}$ }
\bigskip\medskip
\centerline{\sl ${}^a$ Institute for Theoretical Physics, University of
California, Santa Barbara, CA
93106-4030}
\medskip
\centerline{\sl ${}^b$ Department of Physics and Astronomy, University
of
Massachusetts,
Amherst, MA 01003-4525}
\medskip
\centerline{\sl ${}^c$ Physics Department, Syracuse University,
Syracuse, NY 13244-1130}
\medskip
\centerline{\sl ${}^d$ Centro de Estudios Cient\'{\i}ficos de Santiago,
Casilla
16433, Santiago 9, Chile}}
\bigskip
\bigskip

\centerline{\bf Abstract}
\bigskip
We study fully localized BPS brane
solutions in classical supergravity
using a perturbative approach to the coupled Born-Infeld/bulk supergravity
system.  We derive first order bulk supergravity fields
for world-volume solitons corresponding to
intersecting M2-branes and to a fundamental string ending on a D3-brane.
One interesting feature is the appearance of certain off-diagonal metric
components and corresponding components of the gauge potentials.
Making use of a supersymmetric ansatz for the exact fields, we
formulate a perturbative expansion which applies to M2$\perp$M2 $(0)$,
M5$\perp$M5 $(3)$ and D$p$$\perp$D$p$ $(p-2)$ intersections.
We find that perturbation theory qualitatively distinguishes between
certain of these
cases: perturbation theory breaks down at second order
for intersecting M2-branes and D$p$-branes with $p\le 3$
while it is well behaved, at least to this order, for the remaining
cases.  This indicates that the behavior of the full non-linear
intersecting Dp-brane solutions may be qualitatively different for $p \le 3$
than for $p \ge 4$, and that fully localized asymptotically flat
solutions for $p \le 3$ may not exist.  We discuss the consistency of these
results with world-volume field theory properties.
\bigskip
\Date{May, 1999}

\vfill
\eject
%\draftmode
\newsec{Introduction}
The fact that supersymmetric
gauge theories describe the low energy dynamics of branes in string theory has
yielded many important insights, some of which are reviewed in
\refs{\giveon,\karch}.
The most interesting physics arises when supersymmetry is further broken by
brane intersections.
For example, Witten \witten\ (following earlier work in \hanany) has shown
how to
recover the
results of Seiberg and Witten
 \seiberg\ on ${\cal N}=2$ gauge theories in four dimensions via
intersecting M5-branes in eleven dimensions.  The smooth complex curve
describing the
M5-brane intersection in this construction
provides a geometric realization of the Seiberg-Witten curve,
describing the renormalization group flow
of the Yang-Mills coupling.

The constructions referred to above involve branes embedded in flat space.
One expects that useful complementary
information could be obtained from the curved space descriptions of these same
systems.
Delocalized solutions for intersecting branes, in which the harmonic function
associated with each brane is
smeared out over the directions parallel to the other branes,
have been known
for some time (see
{\it e.g.}
\refs{\aaz\baz\caz\daz\eaz\faz\gaz\haz\iaz\jaz\kaz\laz\maz\naz\nazz\nazw\oaz
\paz\qaz {--}\raz}, and
especially
\refs{\gauntlett\stelle{--}\youmbh} for reviews).  These solutions are
useful for many
purposes, such as black hole entropy counting constructions, in which the
delocalized directions are
compactified on a torus. However, the smearing
wipes out much of the interesting physics.

Despite a good deal of effort, spacetimes describing fully localized BPS
brane intersections
have proved quite difficult to find.
In fact, it has been shown \refs{\don,\amdon}
that certain intersections are necessarily delocalized. These results, however,
were limited to either intersections in which one brane is fully contained
within another, or
to ``partially'' localized intersections in which one of the branes is
smeared over the directions parallel to the other brane.
In this paper we will consider the existence of
supergravity solutions for a class of fully localized intersections.
By fully localized,
we mean that the spacetime fields have nontrivial dependence on the coordinates
`along' either brane, and that the sources are appropriate delta functions.
Unlike the cases considered
in \refs{\don,\amdon}, we are now faced with a
nonlinear set of field equations to
solve. Consequently, our analysis will be restricted to a
weak field perturbative
expansion. We start with M2-branes, and solve  perturbatively for the
spacetime fields due to
a non-planar M2-brane source; namely, one for which the M2-brane surface is
a certain
holomorphic curve associated with orthogonal intersecting branes.
However, we find that the second order term
in the perturbation
series diverges.  Intersections of D2-branes (D$2\perp$D$2(0)$) and
D3-branes (D$3\perp$D$3(1)$) behave similarly.  On the other hand,
in the case of M5-branes or D$p$ branes with $p \ge 4$, the second order
term is finite.
One can interpet the second order results as supporting the gauge
theory arguments of \amdon\
about the existence/nonexistence of fully localized BPS intersecting
D-brane solutions.
Possibly the divergence in the perturbative expansion indicates
a more general result that there are no fully localized,
non-planar, static, gravitating BPS M2-branes or D$p$-branes for $p \le 3$.

The question of existence probes an interesting
aspect of the gauge theory description of brane dynamics.
A version of the AdS/CFT limit \refs{\malda,\itzakhi} implies \amdon\ that the
near horizon properties of intersecting D-brane spacetimes have a dual gauge
theory description.
The scale over which brane intersections are delocalized in classical
supergravity
turns out to be
dual to the quantum fluctuations of a massless modulus field in the gauge
theory.
Complete delocalization occurs when these fluctuations become large due to
infra-red
effects.  This in turn is
determined by the dimensionality of the
intersection.
The gauge theory analysis in \amdon\ accounts for the supergravity
results that, for example,
D$p$-branes cannot be localized within D$(p+4)$-branes
for $p=0$ or $p=1$, while they may be
localized for $p\ge2$.

As a starting point for the present analysis, consider the delocalized
solution for
a pair of M2-branes, one in the $(t,1,2)$ plane and one in the
$(t,3,4)$ plane, intersecting at the origin.
The spacetime fields are
\eqn\mbranes{\eqalign{
ds^2=&-(f_1f_2)^{-2/3}dt^2 + f_1^{-2/3}f_2^{1/3}(dx_1^2+dx_2^2)
+f_1^{1/3}f_2^{-2/3}(dx_3^2+dx_4^2) \cr & + (f_1f_2)^{1/3}(dx_5^2 +\dots
+dx_{10}^2)\cr
A_{t12}=& f_1^{-1},\qquad A_{t34}=f_2^{-1},\qquad
f_i=f_i(x_5,\dots,x_{10}),\qquad \nabla_\perp^2 f_i=0\cr},}
where $\nabla_\perp^2=\partial_5^2+\dots+\partial_{10}^2$.
One can try to find localized intersections by starting with an ansatz of
the form \mbranes\
and allowing the functions $f_1$ and $f_2$ to depend on all the spatial
coordinates, {\it i.e.}, not only the directions
$x_5,\dots,x_{10}$ transverse to both branes, but also $x_1,x_2,x_3$ and $x_4$.
However, the equations of motion turn out to require that at least one of
the branes
remains delocalized
\refs{\LP\yang\youm{--}\fayya}. If for example the M2-brane in the (t34)
plane is to be
localized, we must have translation invariance in the $x^3, x^4$
directions, ie, the
(t12) brane cannot be localized. Furthermore
the supergravity equations then reduce to
\eqn\partially{\nabla_\perp^2 f_1=0,\qquad \nabla_\perp^2
f_2+f_1(\partial_1^2+\partial_2^2)f_2=0 .}
Given a solution to the first equation for $f_1$, the second equation for
$f_2$ is then linear\foot{The partially localized intersections studied in
\refs{\don,\amdon}
 are solutions to equations having
this form, but with different numbers of relative transverse and overall
transverse directions.
The relative transverse directions are taken to be compact and $f_2$ is
expanded in Fourier modes.
Solutions for which $f_2$ is localized
can always be found when the two branes are separated in the transverse
directions. However, as the transverse separation
is taken to zero, the Fourier modes with non-zero
wave number are driven to zero unless the number $d$
of overall transverse directions satisfies $d \le 3$.}.

It seems clear that in order to find fully localized
intersections, one must consider a wider class of spacetimes.
Some physical input is then necessary to determine an appropriate
generalization of  the
diagonal ansatz.
Our strategy is quite simple.  The key ingredients in determining the
weak field limit of a given brane
intersection are appropriate source terms for the field equations.
These are provided by coupling the bulk supergravity
fields to brane sources via the Born-Infeld effective action.
Now, since M2-branes have no world-volume gauge fields,
the term `Born-Infeld dynamics' may seem inappropriate.  However,
under dimensional reduction what we refer to here as the Born-Infeld
action of M2-branes reduces to the familiar Born-Infeld action for D-branes
and it is convenient to use the same term for both systems.

Smooth world-volume solitons, sometimes called ``BIons'',
describing certain non-planar branes in a background flat spacetime have been
studied, beginning with the work of \refs{\witten,\callan\gibbons{--}\howe}.
Following the literature, we will refer to these non-planar branes as
``intersecting".
These solitonic configurations are appropriate
sources for the bulk field equations linearized around flat
space.  In section 2, we work out the linearized fields for the
BIonic source describing intersecting
M2-branes.
We will see from the linearized analysis that certain off-diagonal terms in
the metric, as
well as certain additional components of the gauge field, are generated by
BIonic sources.

Any BIon generates a solution to the bulk field equations
linearized about flat space.  Distinctions between different intersecting
brane configurations, however, arise when we look at higher orders in
the weak field expansion.  Carrying out a perturbative expansion based
directly on the field equations
would be a tedious task. Fortunately, the supersymmetry of the BIonic
sources leads to considerable
simplifications. Recently, Fayyazuddin and Smith \fayya\ have presented an
ansatz for the
spacetime fields of fully localized intersecting M5-branes, which preserves
half of the
$32$ supersymmetries of $D=11$ supergravity\foot{However, the
actual intersecting solutions displayed in \fayya\ are diagonal and
describe one localized and one delocalized brane.}.
The remaining undetermined function in the ansatz
satisfies a nonlinear set of equations.

It is straightforward
to alter the form of the ansatz in \fayya\ to
obtain an appropriate ansatz for M2-brane intersections and this is done in
section 3.  The altered ansatz does in fact guarrantee the existence of Killing
spinors appropriate to intersecting M2-branes.
A useful consistency check is that the ansatz matches the
linearized solutions of
section 2.
By dimensional reduction and T-duality, the same
is true for solutions of D$p$-branes intersecting D$p$-branes on
$p-2$ dimensional spatial manifolds in $D=10$ type II supergravities.
We use this structure to investigate the second order perturbations to the
bulk supergravity fields.
Although they are finite for larger branes,
for intersecting M2-, D2-, and D3-branes, we find that the second
order perturbations diverge at every point in spacetime as we take a delta
function
limit of smooth sources to represent fully localized branes.
Section
3 shows that this holds for a simple `crossed-brane' configuration, while
the more complicated calculations associated with holomorphic curve
brane configurations are presented in appendix A.
As will be discussed in section  4,
this meshes well with arguments of \amdon\ based on the low-energy field theory
on the D-branes and is likely to be connected with interesting properties
of full
non-linear solutions.

Lastly, appendix B considers
the weak coupling limit of a fundamental string
ending on a D3-brane.  In this case,
we do not yet have an ansatz for the full non-linear space-time
fields, but we hope that our first order results
will help to motivate such an ansatz. The weak field results do show that the
Born-Infeld spike soliton of \refs{\callan\gibbons{--}\howe}\
generates the appropriate NS anti-symmetric
tensor field to be identified with a fundamental string.

\newsec{M2-Brane Intersections}
We begin by studying the weak field limit of a pair of M2-branes
intersecting at
a point.
We take the action to be given by $S=S_{bulk}+S_{brane}$, where
$S_{bulk}$ is
the $D=11$ supergravity
action and $S_{brane}$ is the Born-Infeld action for an M2-brane
embedded in
curved spacetime.
The bosonic parts of these are given by
\eqn\bulkaction{S_{bulk}={1\over l_{pl}^9}
\int d^{11}x\left\{ \sqrt{-g}\left(R-{1\over 12}F^2\right)
+{2\over
(72)^2}\epsilon^{\mu_1\dots\mu_{11}}F_{\mu_1\mu_2\mu_3\mu_4}
F_{\mu_5\mu_6\mu_7\mu_8}
A_{\mu_9\mu_{10}\mu_{11}}\right\}}
\eqn\BIaction{S_{brane}=-T\int d^3\xi \left\{
\sqrt{-det\, G} - {1\over 6}\epsilon^{abc}(\partial_aX^\mu)
(\partial_bX^\nu)(\partial_cX^\rho)A_{\mu\nu\rho}\right\} .}
Here $G_{ab}=(\partial_aX^\mu)(\partial_bX^\nu)g_{\mu\nu}$ is the
induced metric
on the M2-brane
world volume, $g_{\mu\nu}$ and $A_{\mu\nu\rho}$ are the spacetime metric
and
gauge field
respectively and the M2-brane tension $T$ is related to the $D=11$
Planck length
by $T=1/l_{lp}^3$.
An M2-brane configuration $X^\mu(\xi)$ carries stress-energy
$T_{brane}^{\mu\nu}$
given by
\eqn\stress{\eqalign{
T_{brane}^{\mu\nu}(x) &\equiv {1\over \sqrt{-g}}{\delta S_{brane}\over
\delta
g_{\mu\nu}(x)}\cr
&= -{T\over 2\sqrt{-g}}\int d^3\xi \sqrt{-det\,
G}\,G^{ab}(\partial_aX^\mu)(\partial_bX^\nu)\, \delta^{11}(x-X(\xi))
.\cr}}
and current density for the antisymmetric tensor gauge field
$J_{brane}^{\mu\nu\rho}$ given by
\eqn\current{\eqalign{
J_{brane}^{\mu\nu\rho}(x)&\equiv {1\over \sqrt{-g}}{\delta
S_{brane}\over \delta
A_{\mu\nu\rho}(x)}\cr
&= {T\over \sqrt{-g}}\int d^3\xi \,\epsilon^{abc}(\partial_aX^\mu)
(\partial_bX^\nu)(\partial_cX^\rho)\, \delta^{11}(x-X(\xi)) .\cr}}
These are conserved if the M2-brane equations of motion are satisfied.

In a flat and empty background,
intersecting M2-branes can be described by a BPS soliton solution to the
world-volume equations of motion \witten,\callan\ with two separate
asymptotic
regions
\eqn\soliton{
(X^1+iX^2)(X^3+iX^4)=\alpha_0^2,\qquad  X^5=\dots =X^{10}=0.}
For $\alpha_0^2=0$ this describes a pair of orthogonally intersecting
planes in
the
$(1,2)$ and $(3,4)$ planes.  For $\alpha_0^2\ne 0$ the
intersection region is smoothed out this scale and $|\alpha_0|$
is the size of the ``neck'' where the
two branes join.  The Born-Infeld equations of motion in the flat
background
are satisfied for any $\alpha_0$ in the complex plane.
Now, the Born-Infeld effective action is an approximation to the
low-energy
brane dynamics valid when certain derivatives are small.
For large $\alpha_0$,
all curvatures of the brane are small and the
Born-Infeld description will therefore be accurate.
Furthermore, in \LT\ a related intersection of D3-branes
and fundamental strings was studied in
which, due to supersymmetry, the Born-Infeld description could be shown
to be
exact.
In the present case, we again expect that our Born-Infeld description of
the
brane
dynamics is exact for all values of $\alpha_0$.

Note that the parameter $\alpha_0$ has nothing to do with the charges of
the
branes;
the symmetry of the holomorphic curve guarantees that we have the same
number of branes in each of the two planes.  A similar parameter occurs
in
D-brane intersections of the form D$p\perp$D$p(p-2)$
and, in that case, corresponds to a modulus in the low
energy field theory on the branes.

Let us choose
coordinates $\xi^{0,1,2}=X^{0,1,2}$ on the M2-brane world-volume.  Note
that
this
choice introduces an asymmetry between the two asymptotic regions.
The brane stress tensor
$T_{brane}^{\mu\nu}$
evaluated in
a flat background then has diagonal components
\eqn\stressdiag{\eqalign{T^{00}(x)&={T\over 2}\left(1+{\alpha_0^4\over
R^4}\right )\deight ,\qquad
T^{11}(x)=T^{22}(x)=-{T\over 2}\deight,  \cr
T^{33}(x)&=T^{44}(x)=-{\alpha_0^4T\over 2R^4}\,\deight,}}
and also nonzero off-diagonal components
\eqn\stressoff{\eqalign{
T^{13}(x)&=T^{24}(x)={\alpha_0^2 T (x_1^2-x_2^2)\over 2 R^4}\deight, \cr
T^{23}(x)&=-T^{14}(x)={2\alpha_0^2 T x_1x_2\over 2R^4}\deight,\cr}}
where $R^2=x_1^2+x_2^2$ and $\delta^8(x
-x_0)\equiv\delta(x_3-{\alpha_0^2
x_1\over x_1^2+x_2^2})\,
\delta(x_4+{\alpha_0^2 x_2\over
x_1^2+x_2^2})\delta(x_5)\dots\delta(x_{10})$.
The non-zero gauge current density components are
\eqn\currentcomp{\eqalign{
J^{012}& =T\deight,\qquad J^{034}={\alpha_0^4 T\over R^4}\deight,\cr
J^{013}&= J^{024}= -{2\alpha_0^2 Tx_1x_2\over R^4}\deight, \cr
J^{014}& =- J^{023}= -{\alpha_0^2T(x_1^2-x_2^2)\over R^4}\deight.\cr}}
We solve the linearized Einstein equation in the standard way.
Let $g_{\mu\nu}=\eta_{\mu\nu}+h_{\mu\nu}$, define $\gamma_{\mu\nu}=
h_{\mu\nu}-{1\over 2}\eta_{\mu\nu}h,$
and choose Lorentz gauge $\partial^\rho\gamma_{\rho\mu}=0$.
The linearized Einstein equation then reduces to
$ l_{pl}^{-9}
\partial^\rho\partial_\rho\gamma_{\mu\nu}=-2T^{brane}_{\mu\nu}$.
We find that the solution is
\eqn\solution{\eqalign{
ds^2=&-(1-{2\over 3}(f_1 +f_2))dt^2 + (1-{2\over 3}f_1+{1\over
3}f_2)(dx_1^2+dx_2^2)
+ (1+{1\over 3}f_1-{2\over 3}f_2)(dx_3^2+dx_4^2)\cr &
 +(1+{1\over 3}(f_1+f_2))\delta^{kl}dx_kdx_l+ 2\phi\,(dx_1dx_3+dx_2dx_4)
+
2\psi\,(dx_2dx_3-dx_1dx_4), \cr}}
where the four functions $f_1(x)$, $f_2(x)$, $\phi(x)$ and
$\psi(x)$ satisfy the flat, spatial, ten-dimensional Laplace equation
with
different source
terms,
\eqn\laplace{\eqalign{
&\nabla^2f_1= -l_{pl}^6\delta^8(x-x_0),\qquad
\nabla^2f_2= -{\alpha^4l_{pl}^6\over R^4}\delta^8(x-x_0), \cr
& \nabla^2\phi =- {\alpha^2l_{pl}^6(x_1^2-x_2^2)\over
R^4}\delta^8(x-x_0),\qquad
\nabla^2\psi = -{2\alpha^2l_{pl}^6x_1x_2\over R^4}\delta^8(x-x_0).\cr}}
Similarly, the linearized gauge field is given by
\eqn\gauge{\eqalign{
A=&-f_1dx^0\wedge dx^1\wedge dx^2 - f_2dx^0\wedge dx^3\wedge dx^4
+\psi\,(dx^0\wedge dx^1\wedge dx^3+dx^0\wedge dx^2\wedge dx^4)\cr &
+\phi\,(dx^0\wedge dx^1\wedge dx^4-dx^0\wedge dx^2\wedge dx^3).\cr}}
Integral expressions for the functions $f_1(x)$, $f_2(x)$, $\phi(x)$ and
$\psi(x)$ are then
easily obtained using the ten-dimensional Green's function.  The
resulting
expression for {\it e.g.}
$f_1(x)$ is given by
\eqn\fone{\eqalign{
f_1(x)= \lambda  l_{pl}^6 \int d^2y &
\left\{(x_1-y_1)^2+(x_2-y_2)^2+(x_3-\alpha^2y_1/
(y_1^2+y_2^2))^2 \right .\cr & \left .
+(x_4+\alpha^2y_2/(y_1^2+y_2^2))^2+x
_kx^k\right\}^{-4},\cr}}
where $\lambda=1/8\omega_9$ and $\omega_9$ is the volume of the unit
$9$-sphere.
This integral, like the similar ones for $f_2$, $\phi$ and $\psi$,
cannot be
simply evaluated
analytically.  However, the integral expressions can easily be
manipulated to
show that the following
relations, necessary for the Lorentz gauge conditions to hold, are
satisfied
\eqn\relations{\eqalign{
&\partial_1f_1=\partial_3\phi-\partial_4\psi,\qquad
\partial_2f_1=\partial_4\phi+\partial_3\psi\cr
&\partial_3f_2=\partial_1\phi+\partial_2\psi,\qquad
\partial_4f_2=\partial_2\phi-\partial_1\psi.\cr}}

\newsec{Nonlinear Intersecting M2-branes}
In section 2, we saw that it was straightforward to derive the first
order
fields for a localized
intersection of M2-branes. This raises the question of whether our
analysis can
be
extended to a full perturbation scheme.  Here, we study the second order
perturbations, which already show surprising results.

To make the analysis tractable, we first find a compact formulation of
the full
non-linear
intersecting M2-brane problem. Our approach is based on the recent work
of
Fayyazuddin
and Smith \fayya\ on localized M5-brane intersections, in which a
supersymmetric
ansatz is
given for the spacetime fields.  We start by adapting the results of
\fayya\ to
the M2-brane
case. The ansatz in this case, as in \fayya, depends on a single unknown
function
which must satisfy a non-linear partial differential equation.
The spacetime fields found in the last section, because they are
determined by a
BPS source,
give a leading order solution to these nonlinear equations in weak field
perturbation theory.

\subsec{The Non-perturbative Ansatz}
Fayyazudin and Smith \fayya\ have given an ansatz for the spacetime
fields of a pair of M5-branes intersecting on a 3-brane.  In this case,
as with M2-branes intersecting at a point,
each brane has two spatial dimensions not shared by the other.
The ansatz in \fayya\ is built around
a K\"ahler metric on this four-dimensional relative transverse space,
with the unknown function
being the K\"ahler potential. Define
complex coordinates $s=x^1+ix^2$ and $v=x^3+ix^4$ on the
relative transverse space.
The ansatz for intersecting M2-branes, analogous to that in \fayya, is
then
given by
\eqn\fsmetric{
ds^2 = - H^{-2/3} dt^2 + 2 H^{-2/3} g_{m \overline n} dz^m dz^{\overline
n} + H^{1/3} \delta_{\gamma \sigma} dx^\gamma dx^\sigma,}
where $\gamma,\sigma=5,\dots, 10$, $z^m$ ranges over $s,v$ and $g_{m
\overline
n}$ is a K\"ahler
metric on the associated four-space; i.e.,
we may introduce the potential $K$
such that $g_{m \overline n} = \partial_m \partial_{\overline n} K$.
The function $H$ is related to the
`determinant' $g$ of $g_{m \overline n}$:
\eqn\determinant{
 H = 4g = 4(g_{v \overline v} g_{s \overline s} - g_{v \overline s}
g_{s \overline v}).}
Taking the three-form gauge potential $A$ to be  related to the metric
through
\eqn\A{A_{0 m \overline n} ={1\over 2} i H^{-1} g_{m \overline n}, }
and calculating the supersymmetry variations of the fields shows
that this
ansatz guarantees the existence of Killing spinors $\eta$ satisfying
the projection condition
\eqn\proj{\Gamma_{0 m \overline n} \ \eta = i H^{-1} g_{m \overline n} \
\eta.}
It therefore yields a
supersymmetric solution of 11-dimensional supergravity when the
equations of motion for the gauge field are satisfied.    One can
show that these reduce
to the same nonlinear equation for the K\"ahler potential $K$
found for M5-brane intersections in \fayya,
\eqn\fssource{
{1 \over 2}  \partial_m \partial_{\overline n} (8g(K) + \partial_\gamma
\partial_\gamma K) = J_{m \overline n}, }
where in this case the source $J_{m \overline n}$ is related to the
3-form
current $J^{\mu\nu\rho}_{brane}$ defined in \current\ by
\eqn\rel {J_{m \overline n} =  {i \over 2} \epsilon_{m m_1}
\epsilon_{\overline n
\overline n_1} J_{brane}^{0 m_1 \overline n_1}\sqrt{-\det g_{11}}, }
where $\det g_{11}$ is the determinant of the full eleven-dimensional
metric.  The $i$ in \rel\ guarrantees that $J_{m \overline n}$ is
Hermitian.
Note that,
because the `determinant' $g = H/4$ is a density, rather than a
scalar, equation \fssource\ is not a tensor equation.
Consequently, the form of \fssource\ is
invariant only under holomorphic changes of the coordinates $z^m$ for
which
the Jacobian is the identity.

Following our general strategy, the source $J_{m \overline n}$ should
be consistent with the coupled bulk/Born-Infeld dynamics.  Introducing
a complex spatial coordinate $\xi=\xi^1+i\xi^2$ on the M2-brane world
volume,
one can check that the BI equations of motion are satisfied for any
static
holomorphic configuration
$X^\gamma =0$, $X^0 = \xi^0$, $X^m = X^m(\xi)$, $X^{\overline n}
= X^{\overline n}(\overline \xi)$,
whenever the bulk fields have the form described by \fsmetric\ and \A .
One technical difficulty is that for a single brane with a delta
function
source,
as in section 2, the metric and three-form potential diverge at the
source and
consequently the Born-Infeld equations of motion are not well defined.
In order to deal with this, one may introduce
a `fluid' or `dust' of M2-branes which provides a smooth source at which
the
bulk
fields need not diverge.  Given the non-linearities, the existence of
smooth solutions is non-trivial even for smooth sources.  However, if we
assume that they do in fact exist for arbitrary smooth sources,
one may consider a limit
in which the dust density approximates a $\delta$ function.  In this
sense,
any any holomorphic embedding of the M2-brane is a consistent
source for the full coupled non-linear problem.  In particular, since
an arbitrary smooth source does lead to smooth bulk fields at first
order
in perturbation theory, any holomorphic embedding of the M2-brane is
a consistent source at second order in perturbation theory.

Before studying the weak field perturbation expansion of equation
\fssource\ for
the
K\"ahler potential, we introduce a new set of holomorphic coordinates
for the
relative transverse space $(s,v)$, which will be useful in keeping
the calculations compact.
In complex coordinates the world-volume of the M2-brane source is given
by
the holomorphic curve $sv = \alpha_0^2$.
This is most easily described
by making a holomorphic change of coordinates with unit Jacobian:
\eqn\newcoords{ \alpha = \sqrt{sv},\qquad \beta = \sqrt{sv} \ln{s/v}.}
Translated into the present notation, the intersecting M2-brane
holomorphic curve of section 2 yields a source of the form
\eqn\Jaa {J_{\alpha \overline \alpha } = {q \over 2}
\delta^{(2)} (\alpha - \alpha_0)
\delta^{(6)}(x_\perp)}
with $J_{\alpha \overline \beta}$, $J_{\beta \overline \alpha}$
and $J_{\beta \overline \beta}$ vanishing and $x_\perp$
representing $x^\gamma$ for $\gamma = 1$ to $6$.
Here $q$ is a charge describing the number of branes that
are present and $\alpha_0$ is the parameter describing the ``neck'' of
the
holomorphic curve.  Note that such a holomorphic source satisfies the
obvious integrability condition for the existence of a solution to
\fssource : there is a potential $J = {q \over 4 \pi} \ln |\alpha -
\alpha_0|
\delta^{(6)}(x_\perp)$ such that $J_{m \overline n}
= \partial_m \partial_{\overline n} J$.
In terms of the $(\alpha,\beta)$ coordinates, the field equations and
source
are just as for a flat brane, which will simplify the calculations
below.
Note however, that what makes the problem nontrivial in the coordinates
$(\alpha,\beta)$
are the boundary conditions.
These are determined by the fact that the asymptotic metric takes
the standard Cartesian form
in terms of the original $s,v$ coordinates, and that the  $s,v$
coordinates
are to range over (exactly)
the complex plane.  The result is that the coordinate
$\beta$ ranges only over a strip, and that the asymptotic form of the
metric
is complicated in terms of $\alpha$ and $\beta$.  Thus, it is nontrivial
to construct the exact solution.

However, for the purposes of this paragraph only,
let us make the assumption that the boundary conditions
at infinity are not important near the source.  In this case, the
standard
flat-brane solution holds
(approximately) in this region.
One obtains a solution in which, as usual, the `source' at
$\alpha = \alpha_0$ is replaced by a horizon through which the solution
may be smoothly continued.  Thus, if the boundary conditions are indeed
unimportant near the source, we should in the end obtain a solution of
the sourceless 11-dimensional supergravity equations.

\subsec{Perturbation Expansion}
As noted in \fayya, the nonlinear equation \fssource\ for the Kahler
potential
$K$ can be
solved using a weak field expansion.
Expand the K\"ahler potential as $K = \sum_{ n \ge 0} K^{(n)}$, where
$K^{(n)}$
is
proportional to $q^n$,
and also introduce $g^{(n)}_{n \overline m}
= \partial_n \partial_{\overline m} K^{(n)}$.  We want to perturb
around flat spacetime, so the
zeroth order K\"ahler metric is $g^{(0)}_{m \overline n} = \delta_{m
\overline
n}$
(with $\delta_{s \overline s} = {1 \over 2}$), which follows from the
zeroth
order
K\"ahler potential
$K^{(0)} = {{s \overline s + v \overline v} \over 2}$.    Since we
perturb
around
flat spacetime, the asymptotic boundary conditions will play a central
role.

The nonlinear equation for the K\"ahler potential
\fssource\ is the same for both the
M2-brane intersections considered here and the M5-brane intersections
studied in
\fayya.
Solutions for intersecting D2-branes can be constructed by considering
the setup
for M2-branes,
taking the source to be independent of $x_{10}$ ({\it i.e.}, smearing
the branes
along this direction), and using dimensional reduction.
Further smearing of the
source can create additional symmetry directions, and we can then use
classical
T-duality of the supergravity and Born-Infeld theories to construct  a
fully localized D$p$$\perp$D$p\,(p-2)$
solution in type II supergravity coupled to an appropriate brane
source.  Thus,
by
letting the index $\gamma$ range over an appropriate number ($d=7-p$) of
transverse directions, equation  \fssource\ in fact describes
intersecting solutions
of the form D$p$$\perp$D$p\,(p-2)$.  However, as we perturb around flat space and
impose asymptotically flat boundary conditions in the $d$ dimensional transverse space, 
we will only analyze the cases
with $d \ge 3$ in detail below (i.e., D$p$$\perp$D$p\,(p-2)$ with $p \le 4$ or
intersecting M2- or M5-branes).

Given the form of the zeroth order fields,
the first order terms in \fssource\ combine to give
\eqn \first {
{{1} \over 2} \nabla^2 ( g^{(1)}_{m \overline n}) = J_{m \overline n}, }
where $\nabla^2$ here denotes the $(d+4)$-dimensional flat Laplacian
in the $4$ relative transverse coordinates $s, \overline s, v, \overline
v$ and
the
$d$ overall transverse coordinates $x^\gamma$,
$\nabla^2 = 4 \partial_s \partial_{\overline s} + 4 \partial_v
\partial_{\overline v} + \partial_\gamma \partial_\gamma$.
Let us introduce
the notation $[s] = -1 = [\overline s], [v] = +1 = [\overline v]$.
Then,
with $m$ ($\overline n$) ranging over $s,v$ ($\overline s, \overline
v$),
all the source components of equation \currentcomp\ assemble using \rel\
into
the compact form
\eqn \moresource { J_{m \overline n} = e^{[m] \beta /2 \alpha}
e^{[\overline n]  \overline \beta /2 \overline \alpha} {q \over 2}
\delta^{(2)}(\alpha - \alpha_0) \delta^{(d)}(x_\perp),}
and the components of the first order K\"ahler metric are given by
\eqn\gone{g^{(1)}_{m \overline n}  = {{-q} \over {(d+2) \omega_{d+3}}}
\int d^2 \beta'
{{e^{[m] \beta'/ 2\alpha_0} e^{[\overline n] \overline \beta'/2\overline
\alpha_0} }
\over
( \sum_{\gamma = 1}^d x^\gamma x^\gamma
+ | s - \alpha_0 e^{-\beta'/2\alpha_0} |^2 +
|v - \alpha_0 e^{\beta'/2\alpha_0} |^2)^{(d+2)/2}},}
where $\omega_{d+3}$ is again the volume of the unit (d+3)-sphere.
These results are a more compact form of those given in equation
\solution\ in section 2 for $d=6$.

The sources $J_{m\bar n}$ do not
depend on the background metric. Therefore,
the right hand side of \fssource\ only has contributions at
first order. Continuing to the expansion of \fssource\ we find that the
terms
of order $j$ satisfy
\eqn\Kn{\partial_m \partial_{\overline n} \left( \nabla^2 K^{(j)}
+ 8 \sum_{ 1 \le k \le j-1} \left(
g^{(k)}_{s \overline s} g^{(j-k)}_{v \overline v} -
g^{(k)}_{v \overline s} g^{(j-k)}_{s \overline v} \right) \right) = 0,}
for $j >1$. Boundary conditions at infinity for localized branes imply
that the
quantity in
parenthesis must vanish. Hence, the higher order terms in $K$ satisfy
a flat ten dimensional Laplace equation with sources given by
products of lower order terms and are
given formally by the integrals
\eqn\Knsol{
K^{(j)} (x_0) = {4 \over { (d+2) \omega_{d+3}}} \int_{}^{} d^{10} x
{\left(g^{(k)}_{s \overline s} g^{(j-k)}_{v \overline v} - g^{(k)}_{v
\overline s} g^{(j-k)}_{s \overline v} \right )
\over {|x_0 -x|^{d+2}}} ,}
where the notation $x_0,x$ includes the complex coordinates $s,v$ as
well
as the transverse coordinates $x^\gamma$.  When the integral \Knsol\
converges, it gives the unique solution to \Kn\  satisfying the
appropriate
boundary conditions.

\subsec{To Converge, or Not to Converge?}
The important question which needs to be addressed is whether the
integrals \Knsol\ for $K^j$ do in fact converge,
starting with the second order term $j = 2$.
We consider here the limit $\alpha_0 \rightarrow 0$ in which
the smooth intersection degenerates into the singular intersection
of two perpendicular
planes.  Although, due to the large curvature at the intersection, the
Born-Infeld
description of the dynamics is not a priori justified in this limit,
the considerably more complicated calculations for $\alpha_0\ne 0$ lead
to the
same
conclusions.  These calculations are presented in appendix A.
In the $\alpha_0 \rightarrow 0$ limit, the
nonzero source terms in \Jaa\ are given simply by
\eqn\nonzero{J_{s \overline s} = {q \over 2} \delta^2(s) \delta(x),
\qquad
J_{v \overline v} = {q \over 2} \delta^2(v) \delta(x).}
The fact that $J_{s \overline v},J_{v\bar s} \rightarrow
0$ even at $s=v=0$ can be verified  by integrating
$J_{s \overline v}$ (at finite $\alpha_0$) over any region invariant
under $s \rightarrow e^{i \theta} s$, $v \rightarrow e^{-i\theta} v$.
The first order metric for the crossed-plane
source is just the superposition of the results for flat branes at
$s=0$ and $v=0$.  For example,  the component
$g^{(1)}_{s \overline s}$
is
\eqn \limgone {g^{(1)}_{s \overline s} =  {{-q \over { (d+2)
\omega_{d+3} (x^\gamma x^\gamma
+ |s|^2)^{d/2}} }},}
with an analogous expression for $g^{(1)}_{v \overline v}$.
The off-diagonal
terms $g_{s \overline v}$ and $g_{v \overline s}$ both vanish.

The integral in \Knsol\ for $K^{(2)}$ then has the form
\eqn \Ktwo{{}K^{(2)} (x_0) = {4 \over { (d+2)^3 \omega_{d+3}^3}}
\int {{q^2} \over {|x_0 -x|^{d+2}}} {{d^{d} x d^2s d^2v}
 \over {(x{}^\gamma x{}^\gamma + |s|{}^2)^{d/2}
(x{}^\sigma x{}^\sigma + |v|{}^2)^{d/2}}} .}
Let us analyze this integral in a small region
near $x^\gamma=s=v= 0$.  In this region, we may
approximate $|x_0 -x|$ by a constant.  Introducing
$\rho^2 = x{}^\gamma x{}^\gamma + |s|{}^2 + |v|{}^2$, the integral over
this small region factors into an integral
over angles and an integral over $\rho$ of the form $\int \rho^{3-d} d
\rho$.
The integral over angles does not vanish as the integrand is strictly
positive.  Thus, when $d \ge 4$, the integral diverges for any $x_0$.
However, for $d = 3$, the integral converges and the second order
perturbation is well-defined.  Although we have not explicitly considered
the cases with $d < 3$, it is clear that the second order perturbation will have
no short distance divergences in those cases.

This calculation suggests that higher order perturbation theory breaks
down when
the number $d$ of overall transverse dimensions satisfies
$d\ge 4$, which includes the M2-brane intersection ($d=6$).
On the other hand, perturbation theory is potentially well-defined for
$d = 3$,
which includes M5-brane intersections.
As will be discussed further in section 4, these results fit well
with both the supergravity results of \amdon\
in similar, but slightly different, situations and with the predictions
of that
work, based on arguments in the D-brane field theory, for supergravity
solutions of the present form.

Since the divergence of second order perturbations may be unexpected,
the reader
may wonder if some subtlety has been passed over through the use of
singular
sources.
To show that such subtleties are under control, we consider below the
same
calculations for smooth sources and study the limit in which the smooth
sources approximate the delta-functions above.

\subsec{Smooth Sources}
Still keeping $\alpha_0=0$,
we smooth the sources according to
\eqn \smoothsource{J_{s \overline s} = {q \over 2}
f_L(|s|^2 + x^\gamma x^\gamma),\qquad
J_{v \overline v} = {q \over 2}
f_L(|v|^2 + x^\gamma x^\gamma) ,}
where $f_L$ is a smooth, non-negative function
which vanishes for
$r>L$ and has unit
normalization; {\it i.e.}, satisfying $\omega_d \int_0^L f(r) r^{d+1} dr
= 1$.
Note that this
smoothing is simple to carry out because of the ``no force''
condition between BPS objects.
With sources smoothed over any scale $L$, solutions
exist at each order of perturbation theory. We want to study
the behavior of solutions as
we take $L\rightarrow 0$. Typically there are many ways to take such
a limit in general
relativity (see {\it e.g.} \geroch). However, the present BPS system
is highly constrained. Fixing the
volume integral of the current components $J_{m\bar n}$ determines the
total
charge via Gauss's law,
$\nabla_\mu F^{\mu\rho\sigma \tau}=J^{\rho\sigma \tau}$. Therefore for
each $L$
the
solution has the same charge in
the above prescription.  Expanding in multipole moments, we see that to
leading
order
in $r^{-1}$, $K^{(1)}$ stays the same
for all $L$.

Symmetry considerations guarantee that the first order fields evaluated
outside
the
dust distribution are identical to those from the delta-function source,
and that $g_{s \overline v}$ and  $g_{v \overline s}$ remain identically
zero.
However, the first order fields are now smooth everywhere, so the
integral
defining $K^{(2)}$ converges.  It therefore gives the correct second
order
perturbation for the smooth source.

Now, consider the limit in which $L\rightarrow 0$ and $f_L$ becomes the
appropriate delta function.  For
any smooth $f_L$ approximating the singular source, we may divide the
integral
for $K^{(2)}$ into an integral over a region outside the support of
$f_L$, and
one over a region inside.  Since the integrand in the outside region
is just the same as in the delta-function case, we have already seen
that, for $d \ge 4$, it grows without bound in the limit.  Now note that
since
$f$ is non-negative,  $g^{(1)}_{s \overline s}$ and $g^{(1)}_{v
\overline v}$
are positive and the source for $K^{(2)}$ is of a definite sign.
Thus, the integral over the region containing the source contributes
to $K^{(2)}$ with the same sign as in the exterior region.  Thus, we
conclude
that for $d \ge 4$,
in the limit in which the smooth source becomes a delta function,
$K^{(2)}$ grows without bound
at each $x_0$.

The effect of this divergence on a physical quantity is somewhat
subtle.  For
example, although the divergence occurs at the same order in $r^{-1}$ as
the
term from $K^{(1)}$ that encodes the total charge, it cannot in fact
effect
the total charge computed at infinity.  This is fixed by charge
conservation,
and the divergence can only appear in $F^{\mu \rho \sigma \tau}$ at
higher
order in $r^{-1}$.

It is useful to note that arguments of the above form apply directly to
the second order metric perturbation
$g^{(2)}_{m \overline n }$, and to the norm $||\partial_t||^2 \sim
H^{-2/3}$
of the timelike Killing field.  The latter is a scalar under coordinate
transformations, so that its divergence shows that the result is not
an artifact of our particular choice of gauge.   Thus, we conclude
that perturbation theory breaks down at
second order for
localized solutions of intersecting M2-, D2-, and D3-branes.  However,
the second order perturbations do exist for localized
 intersecting solutions of larger
branes for which $d \le 3$.

Now, on the one hand, it is no surprise that perturbation theory cannot
construct a full non-linear solution corresponding to a delta function
source.  We expect a full solution to have a horizon, which is a strong
field effect.  Sources may be characterized by a `charge radius'
$r_c \sim q^{1/(d-1)}$, and by a length scale $L$ associated with
the support of $f$.  One expects perturbation theory to be useful
for weak sources with $r_c/L \ll 1$, but not for strong sources with
$r_c
/L \sim 1$ or greater.  Of course, the difference between weak
and strong sources is usually apparent only when one attempts to sum
the perturbation series.  What is interesting about our case is
the explicit divergence of the second order term and the fact that the
behavior is very different for $d \le 3$ than for $d \ge 4$.  Although
we can say nothing definite about the full non-linear solutions, this
strongly suggests that their behavior is qualitatively different
for $d \ge 4$ than for $d \le 3$.  In particular, it is consistent
with the prediction of \amdon{} that fully localized asymptotically
flat solutions should
exist only for $d \le 3$. In appendix A, we show that the same behavior
holds
for
$\alpha_0\ne 0$.

\newsec{Discussion}
In this work we have explicitly constructed the first order perturbative
bulk supergravity fields corresponding to intersecting M$2$-branes.
The corresponding results for a fundamental string ending on a
D$3$-brane appear
in appendix B.  We also
showed that, as one would expect,
any solution of the coupled bulk supergravity/Born-Infeld
system for intersections of the form
M$2$$\perp$M$2\,(0)$ or D$p$$\perp$D$p\,(p-2)$ is controlled by
equations of
the form presented in \fayya\ for M$5$$\perp$M$5\,(3)$.  We used this
structure to analyze the second order perturbations of the bulk
fields.  While these perturbations are finite and small far from
the branes for intersecting M$5$-branes and intersecting
D$p$-branes with $p \ge 4$, the
second order perturbations diverge everywhere in the spacetime
for intersecting M$2$-branes and for intersecting D$p$-branes with
$p = 2,3$.

This result appears to fit well with the predictions
of \amdon\ based on field theory considerations.  That work
started from the observation \don\ that there are no fully localized
solutions for one-branes inside five-branes.  Solutions do
exist when the branes are separated in the transverse direction, but the
one-branes necessarily delocalize as the transverse separation is
removed.  The
limit of
zero separation gives one-branes `smeared' over the five-branes.
It was shown in \amdon\ that similar results hold in a number of other
contexts,
such
as D$(p-4)$-branes parallel to D$p$-branes for $p=3,4$, or
D$p$-branes intersecting smeared D$p$-branes on a $p-2$ surface for $p
\le 3$.
This behavior is in contrast with the situation for larger branes in
which the solutions remain localized as the transverse separation
is removed\foot{The first such localized solutions were
found in \ITY\ in the near-core limit.}.

Similar effects are found in certain `near-horizon' spacetimes.
Therefore, one expects to have a field theory description of this effect
through an analogue of the dualities  described in \malda,\itzakhi.
Understanding the field theory origin of delocalization was the main
goal of
\amdon. Consider first the case of D$(p-4)$-branes parallel to
D$p$-branes.
Since both are associated with a `width' of the D($p-4$) branes in the
directions
along the D$p$-branes, a
natural idea is that the delocalization in classical supergravity is
somehow related to the scale size of the instantons that describe
the smaller branes in the Higgs phase of the D$p$-brane field theory.
In dualities in general, strong field classical effects on one side
are related to strongly quantum mechanical effects on the other.
It turns out that the supergravity delocalization is related
to the quantum fluctuations of the scale size in the field theory.
Fluctuations which would be large due to ultraviolet effects are
suppressed by a
string-scale cutoff, but the fluctuations can still be large due to
infrared
effects.
The relevant field theory lives on the intersection of the two branes
and
delocalization occurs in exactly those cases where this is
0+1 or 1+1 dimensional, for which the infrared effects do indeed make
the
fluctuations
large.  This ``fluctuation-delocalization duality''
correctly predicts both the cases in which the supergravity should
delocalize and the rate at which it does so as the transverse
separation is removed.

Now consider
D$p$-branes intersecting D$p$-branes
on  a surface with  $(p-2)$  spatial dimensions.
Such intersecting
branes are associated with holomorphic
curves $Z_1Z_2 = \alpha^2_0$ in ${\bf C}^2$, where ${\bf C}$ denotes
the complex numbers.
It turns out that $\alpha_0$ is a modulus and is related to
the scale size modulus associated with D$(p-4)$-branes inside
D$p$-branes through T-duality.
Thus, one expects similar behavior in this case, with
delocalization related to the quantum fluctuations of $\alpha_0$.
As little information was
available regarding the classical supergravity solutions for
fully localized intersecting branes, \amdon\ could compare the field
theory only with the classical supergravity solutions in which one
brane was smeared over the worldvolume of the other.  For such cases,
agreement was once again found with regard both to which cases should
delocalize and how fast this should happen as the transverse
separation is removed.

The natural prediction is of course that
a fully localized solution in which two branes are separated
in a transverse direction should also delocalize when this separation
is removed and therefore that fully localized intersecting brane
solutions with D$p$$\perp$D$p\,(p-2)$ should not exist for $p \le 3$.
As a result, one expects that
M-theory solutions with M$2$$\perp$M$2\,(0)$ also should not
exist.
These are just the cases for which we found a divergence of
the second order perturbations of the bulk fields.  Note that
since first order perturbation theory is linear, the lack of a
well-defined
second order perturbation is the natural signature of the non-existence
of fully localized asymptotically flat solutions.

A small subtlety is that one should remember that the field theory
is dual to the supergravity physics only in the near-horizon
region.  As a result, it is not clear just what the field theory
arguments
have to say about the existence of asymptotically flat (as opposed to
near-horizon) supergravity solutions for which the neck size $\alpha_0$
of the supergravity solution is comparable to or larger than the charge
radius $r_c$ of the branes.  For this reason,
\amdon\ could conclude that such solutions
fail to exist only for small $\alpha_0$.   It is interesting
that our perturbative results were qualitatively the same for
all values of $\alpha_0$, but it is not clear to what extent
the existence of full non-linear solutions for large $\alpha_0$
should be reflected in perturbation theory.

Having found that the second order perturbations fail to exist
for $d \ge 4$, it is natural to ask about the higher order perturbations
for the case $d \le 3$.  Do they in fact exist?
This is far from clear.
The source terms for the higher order perturbations are more
complicated,
and there is the potential for subtle cancellations even in the
case $\alpha_0 =0$.   We leave this question for future work.

\medskip
{\centerline{\bf Acknowledgments}}

We would like to thank Patrick Brady, Pablo Laguna, 
Amanda Peet, P. Ramadevi, Larus Thorlacius and, especially,
Jorge Pullin  for for useful discussions.  This work was supported in
part by
National Science Foundation grant No. PHY94-07194.  D.M. and A.G. were
also
supported
in part by NSF grant No.  PHY97-22362 and funds from Syracuse
University. The work of DK and JT is also supported in part by NSF grant
No.
PHY98-01875.

\appendix{A}{Sources with $\alpha_0 \neq 0$}

We now wish to consider the second order perturbation $K^{(2)}$ for the
case
$\alpha_0 \neq 0$.  Again, we will find that $K^{(2)}$ exists only for
$d \leq 3$.   Let us consider the value of $K^{(2)}$ at some point
$x_0 = (x_0^\gamma, s_0, \overline s_0, v_0, \overline v_0)$.
{}From \Knsol{}, this is
\eqn \Ktwo {
K^{(2)} (x_0) = {4 \over {(d+2) \omega_{d+3}}} \int_{}^{} d^{d} x d^2 s
d^2 v
{{
g^{(1)}_{,s' \overline s'} (x) g^{(1)}_{,v' \overline v'}(x) -
g^{(1)}_{,s' \overline v'} g^{(1)}_{,v' \overline s'} }
\over {|x_0 -x|^{(d+2)}}}, }
with $g^{(1)}_{m \overline n}$ given by \gone.
As before, a divergence can only result from integrating over the
singularities
in the first order fields that arise at the location of the source.
Note, in particular, that adding to $g^{(1)}_{m \overline n}$
any smooth function of $x$ with the same large $x$ behavior will not
alter the convergence of the above integral.  This is the strategy
we will invoke below.

If, instead of integrating over the entire $\beta$ strip in \gone{},
we restrict the integration to be over only the
the region $|\beta' - \beta| < 2 \epsilon_0$, then this changes
$g^{(1)}_{m \overline n}$ only by a smooth function of the sort
mentioned above.
In the remaining (small $\beta'-\beta$) region,
it is useful to expand
$e^{[m]\beta'/2 \alpha_0}
e^{[\overline m]\overline \beta'/2 \overline  \alpha_0}$
in powers of $\epsilon/\alpha_0 := (\beta'
- \beta)/2\alpha_0$.  We
write the resulting infinite series as
$e^{[m]\beta/2\alpha_0}
e^{[\overline m]\overline \beta/2 \overline \alpha_0}
P^{(1)}_{m \overline n}$.
The expression
$P^{(1)}_{m \overline n}$
is a series in $\epsilon$, $\overline \epsilon$ with constant
coefficients.

We also expand terms in the denominator
in powers of $\delta/\alpha_0 :=
\alpha/\alpha_0 -1$.
Note that we have
\eqn \ssquare {|s-s'|^2 =  e^{-\beta/2 \alpha_0}e^{-
\overline \beta/2 \overline \alpha_0}
|\delta (1 + \beta/2\alpha) - \epsilon +
O(\delta^2, \delta \epsilon, \epsilon^2)|^2}
and similarly for $|v-v'|^2$.  The singularity of
$|x - x'|$ will be controlled  by
\eqn \capdelta{ \eqalign{ \Delta = &
 x^\gamma x^\gamma + 2 \cosh(\beta/2\alpha_0 + \overline \beta/
2\overline \alpha_0) \epsilon \overline \epsilon \cr
+ &  \left[ (1 + {\beta \over {2 \alpha_0}})(1 + {{\overline
\beta} \over {2 \overline \alpha_0}}) e^{-\beta/2\alpha_0}e^{-
\overline \beta/2 \overline \alpha_0} +
(1 - {\beta \over {2 \alpha_0}})(1 - {{\overline
\beta} \over {2 \overline \alpha_0}}) e^{+\beta/2\alpha_0}e^{+
\overline \beta/2 \overline \alpha_0} \right]
\delta \overline \delta \cr
+ &  \left[ (1 + {\beta \over {2 \alpha_0}})
e^{-\beta/2\alpha_0}e^{-
\overline \beta/2 \overline \alpha_0} -
(1 + {\beta \over {2 \alpha_0}}) e^{+\beta/2\alpha_0}e^{+
\overline \beta/2 \overline \alpha_0} \right]
\overline \epsilon \delta  \cr
+ &  \left[ (1 + {\overline \beta \over {2 \overline  \alpha_0}})
e^{-\beta/2\alpha_0}e^{-
\overline \beta/2 \overline \alpha_0} -
(1 + {\overline \beta \over {2 \overline  \alpha_0}})
e^{+\beta/2\alpha_0}e^{+
\overline \beta/2 \overline \alpha_0} \right]
\epsilon \overline \delta \cr
} ,}
since we may write
\eqn \xsquare {|x-x'|^2 = \Delta + O(\delta^3, \delta^2 \epsilon, \delta
\epsilon^2 , \epsilon^3) = \Delta \left(1 + {{\delta^2} \over \Delta}
O(\delta,\epsilon)+ {{\epsilon^2} \over \Delta} O(\delta,
\epsilon)\right).}
Since, for any $\beta, \overline \beta$ the object $\Delta$ is a
positive
definite quadratic form in $x^\gamma, \delta, \epsilon$, the functions
${{\delta^2} \over \Delta}$,
${{\epsilon^2} \over \Delta}$ are bounded by functions of $\beta,
\overline
\beta$.  Thus, we may write
\eqn \denom {|x-x'|^{-(d+2)} = \Delta^{-(d+2)/2} P^{(2)},}
where
$P^2$ is a series in both
$|{ \epsilon \over \alpha_0}|$,  and $|{\delta \over \alpha_0}|$  whose
coefficients  involve functions of the form
$\epsilon^2/\Delta$ and $\delta^2/\Delta$.
The most important
property of $P^{(2)}$ is that it does not depend on
$m, \overline n$.

Collecting these observations together, we have
\eqn \gsing{ \left(g^{(1)}_{m \overline n} \right)_{sing}
=  - {{  e^{[m] \beta/2 \alpha} e^{[\overline n
] \overline \beta/2 \overline \alpha}
} \over {(d+2) \omega_{d+3}}}
\int_{|\epsilon < \epsilon_0|} {{ d^2 \epsilon \ \ P^{(1)}_{m \overline
n}
P^{(2)}
}  \over
{ \Delta
^{(d+2)/2}}},}

Note that when considering sufficiently high order terms that arise in
the
product $P^{(1)}_{m \overline n} P^{(2)}$,
the integral
over $d^2\beta'$ is nonsingular, even for $x^\gamma=0$, $\alpha =
\alpha'$.
Thus, dropping these terms again changes $g^{(1)}_{m \overline n}$ only
by another smooth function of appropriate decrease at infinity.

Having dropped the terms in $P^{(1)}_{m \overline n}
P^{(2)}$ that are not singular
at $\epsilon = x^\gamma = \delta =0$, let us consider taking $\epsilon_0
\rightarrow \infty$ to remove the restriction on the region of
integration.
The highest remaining terms lead to logarithmic
diverges at large $\epsilon$,
but the other terms remain finite.  Thus, if we add appropriate
counterterms to regulate the logarithmic divergence, taking
$\epsilon_0 \rightarrow  \infty$
changes $g^{(1)}_{m \overline n}$ only
by a bounded function and does not effect the convergence of the
second order perturbations \Ktwo{} to the K\"ahler potential.
The details of treating the large $\epsilon$ logarithms are not
important,
as we will see that the convergence of \Ktwo{} at small $\epsilon$ is
controlled by lower order terms in $P^{(1)}_{m \overline n} P^{(2)}$.

Extending the integration region in this way over the entire complex
$\epsilon$-plane, the integral \gsing{} may be evaluated exactly (see,
for
example, \gr).  The result has the form

\eqn \gsingtwo{ \left(g^{(1)}_{m \overline n} \right)_{sing}
=  - {{  e^{[m] \beta/2 \alpha} e^{[\overline n
] \overline \beta/2 \overline \alpha}
} \over {(d+2) \omega_{d+3}}}
 { {Q_{m \overline n}}  \over
{ \left( x^\gamma x^\gamma + 2 \Omega^2
\delta \overline \delta
\right)^{d/2}}}, }
where $Q_{m \overline n} $
is a polynomial in $|\delta|$
whose coefficients are determined by those of
$P^{(1)}_{n \overline m} P^{(2)}$ and
\eqn \Om {\eqalign {\Omega ^2(\beta, \overline \beta)
&= (1 - \beta/2\alpha_0)(1 - \overline \beta/2\overline
\alpha_0) e^{\beta/2\alpha_0} e^{\overline \beta/2 \overline \alpha_0}
\cr
& +
 (1 + \beta/2\alpha_0)(1 + \overline \beta/2\overline
\alpha_0) e^{-\beta/2\alpha_0} e^{-\overline \beta/2 \overline \alpha_0}
\cr &-{{|
(1 - \beta/2\alpha_0)
e^{\beta/2\alpha_0} e^{\overline \beta/2 \overline \alpha_0} -
(1 + \beta/2\alpha_0)
e^{-\beta/2\alpha_0} e^{-\overline \beta/2 \overline \alpha_0}|^2 }
\over {2 \cosh (\beta/2\alpha_0 + \overline \beta/2 \overline
\alpha_0)}}.}}

We must now see how the various terms in \gsingtwo{} effect the
second order K\"ahler potential \Ktwo{}.  Note that the first order
fields enter quadratically, through the combination $4H^{(1)} =
g^{(1)}_{s \overline s}
g^{(1)}_{v \overline v} - g^{(1)}_{s \overline v}
g^{(1)}_{v \overline s}$.  The singular part of this expression may
be written
\eqn \gsingtwo{ \left(H^{(1)} \right)_{sing}
= {4 \over {(d+2)^2 \omega^2_{d+3} }}
{ {Q_{s \overline s} Q_{v \overline v} - Q_{s \overline v}
Q_{v \overline s} }  \over
{ \left( x^\gamma x^\gamma +  \Omega^2
\delta \overline \delta
\right)^{d/2}}}. }
The effects of a term in $H^{(1)}$ of given order in $|\delta|$
on the second order
perturbation $K^{(2)}$ are straightforward to analyze.  After rescaling
$\delta$  by  $\Omega$,
the $\beta, \overline \beta$ dependence
factors out.  The integral over  $\beta,
\overline \beta$  converges, and
the only integrals remaining to be done are of the form
\eqn \rest { \int {{d^d x d^2 \delta \ |\delta|^k
} \over {(x^2 + |\delta|^2)^{d}}}.}
The convergence of such integrals can be studied by introducing the
radial coordinate $\rho = \sqrt{x^\gamma x^\gamma + | \delta|^2}$.  The
expression \rest{} factors into a convergent angular
integral and a radial integral that converges for $k +
1 \ge d$.

Clearly, the relevant issue is which values of $k$
actually contribute. This is just the question of determining the
smallest
power of $|\delta|$ that appears in the numerator of
$\left(H^{(1)} \right)_{sing}$, which in turn
can be found by studying how the first order fields \gone{} enter
into $\left( H^{(1)} \right)_{sing}$.  Let us first consider terms
of the form \rest{} that arise from the constant term in $P^{(2)}$;
i.e., for the moment take $P^{(2)}=1$.

Note that the first few terms in $P^{(1)}_{m \overline n}$ are
\eqn \Pterms {P^{(1)}_{m \overline n} = 1 + [m] \epsilon/\alpha_0
+ [\overline n]
\overline \epsilon /\overline \alpha_0
+ {{[m]^2} \over 2} \epsilon^2/ \alpha_0^2 +  {{[\overline n]^2} \over
2}
\overline \epsilon^2/\overline \alpha_0^2
+ [m][\overline n] \epsilon \overline \epsilon/ \alpha_0 \overline
\alpha_0 +
O(\epsilon^3/\alpha_0).}
To this same order, taking $P^{(2)}=1$,
the singular part of $H^{(1)}$ is therefore
\eqn \Hexp {\eqalign{ \left( H^{(1)} \right)_{sing}^{P^{(2)}=1} =
{4 \over {(d+2)^2 \omega_{d+3}^2}} \left(
\int {{ P^{(1)}_{s\overline s} d^2 \epsilon} \over{
\Delta^{(d+2)/2} }}
\int {{ P^{(1)}_{v\overline v} d^2 \epsilon} \over{
\Delta^{(d+2)/2} }} -
\int {{ P^{(1)}_{s\overline v} d^2 \epsilon} \over{
\Delta^{(d+2)/2} }}
\int {{ P^{(1)}_{v\overline s} d^2 \epsilon} \over{
\Delta^{(d+2)/2} }} \right) \cr
= {{16} \over {(d+2)^2 \omega_{d+3}^2 \alpha_0 \overline \alpha_0 }}
\left(
\int {{ \epsilon \overline \epsilon d^2 \epsilon} \over{
\Delta^{(d+2)/2} }}
\int {{ d^2 \epsilon} \over{
\Delta^{(d+2)/2} }} -
\int {{ \epsilon d^2 \epsilon} \over{
\Delta^{(d+2)/2} }}
\int {{ \overline \epsilon d^2 \epsilon} \over{
\Delta^{(d+2)/2} }}  \right)  + ...,
}}
as all terms of less than second order cancel out.  The ellipses
above denote terms of higher order.  The important
question is whether the second order terms above also cancel.  It turns
out
that this is not the case.  To see this, write $\Delta$ as
$ A \overline A \epsilon \overline \epsilon + B \epsilon + \overline B
\overline \epsilon + C \overline C
= |(A \epsilon + B/\overline A)|^2 + C \overline C - |B/A|^2$ and
change integration variables to $\omega = A \epsilon + B/\overline A$.
Since $\Delta$ is even in $\omega$, integrals of the form $\int {{
\omega d^2 \omega} \over {\Delta^{(d+2)/2}}}$ vanish.  As a result,
we may write
\eqn \Hexptwo {\eqalign {\left( H^{(1)} \right)_{sing}^{P^{(2)}=1}
= { 16 \over { (d+2)^2 \omega_{d+3}^2 \alpha_0 \overline \alpha_0
|A|^6}}
\int {{ d^2 \omega} \over{
\Delta^{(d+2)/2} }}
\int {{ \omega \overline \omega d^2 \omega} \over{
\Delta^{(d+2)/2} }} \cr = (const)
 { 1 \over { \alpha_0 \overline \alpha_0
|A|^6}}
(x^\gamma x^\gamma + \Omega^2 \delta \overline \delta)^{-2(d-1)} }}
to the same order as in \Hexp.
Note that $A$ depends only on $\beta, \overline \beta$.  Let us define
$\delta_0 = \Omega \delta$.  Then $K^{(2)}$ involves the integral
of the above expression \Hexptwo{} with respect to the measure
$\Omega^{-2} d^2 \beta \  d^2 \delta_0 \ d^dx_\perp$.  The integral
over $\beta, \overline \beta$ converges and clearly gives a result
proportional to $(\delta_0 \overline \delta_0 + x^\gamma
x^\gamma)^{-2(d-1)}$.
We therefore see that the integral over $\delta_0, \overline \delta_0,
x^\gamma$ converges if $d+2 > 2(d-1)$; i.e., for $d \le 3$.  On the
other
hand, for
$d \ge 4$, this contribution to $K^{(2)}$ diverges at every point in
the spacetime.

We have now shown that, for $d \ge 4$,
the terms that arise from the order zero piece
of $P^{(2)}$ cause a divergence in $K^{(2)}$ at order $k=2$ (in the
counting of \rest{}).  To conclude that $K^{(2)}$ is in fact divergent,
we need only show that higher order terms in $P^{(2)}$ cannot cancel
this
divergence.  This is not hard.  Let $P^{(2)(1)}$ be the collection of
first order terms in $P^{(2)}$, proportional to either $|\epsilon|$
or $|\delta|$.  A compensating divergence
could only come from the interaction
of $P^{(2)(1)}$  with a term of order $\epsilon$ or $\overline \epsilon$
in $P^{(1)}_{m \overline n}$. Let $P^{(1)(1)}_{m \overline n}$
denote the first order terms in $P^{(1)}_{m \overline n}$.  Due
to the structure of our system,  $P^{(2)(1)}$ always appears
with either
$P^{(1)(1)}_{s \overline s}+ P^{(1)(1)}_{v \overline v}$
or
$P^{(1)(1)}_{s \overline v}+
P^{(1)(1)}_{v \overline s}$.  However, both of these vanish.
That a higher
order divergence does not arise
from the interaction of
$P^{(2)(1)}$ with the zero order term $P^{(1)(0)}_{m \overline n}$
in $P^{(1)}_{m \overline n}
$ follows from the fact that that
$P^{(1)(0)}_{m \overline n}$  is independent of $m, \overline n$. Thus,
$K^{(2)} (x_0)$ does indeed diverge for all $x_0$ when $d \ge 4$;
i.e., for M2, D2, and D3-branes.

Once again, one may consider replacing the localized intersecting brane
with a smooth dust of branes concentrated in a region of size $L$ in the
transverse directions.
This leads to smooth metric functions $g^{(1)}_{m \overline n}$ which
converge to the localized brane first order fields \gone{} in the
$L \rightarrow 0$ limit.  The analysis proceeds much as in the case
of a delta-function source, but with extra integrals over $\alpha_0$ and
the location of the brane in the $x^\gamma$ directions.  In particular,
$H^{(1)}$ has a similar structure.  Thus, in the limit where the source
becomes a delta function, $K^{(2)}(x_0)$ diverges for all $x_0$ for $d
\ge 4$.
As before, one can also show that $g_{m \overline n}^{(2)}$ and
$||\partial_t ||^2$ diverge as well.

Thus, the second order perturbations are infinite and perturbation
theory
breaks down at second order for $d \ge 4$, though not for $d \ge 3$.
This suggests that the full non-linear localized solutions are
quite different for $d \le 3$ than for $d \ge 4$.  In particular, it is
consistent with the prediction of \amdon{} that localized solutions
should
not exist, at least for small $\alpha_0$.  It is interesting that the
divergence encountered here does not in fact depend on the
value of $\alpha_0$, but it is
not clear if such a feature of the full solutions should be apparent at
this level of analysis.

\appendix{B}{Strings Ending on D3-branes}

In this appendix
we compute weak coupling solutions to the coupled $D=10$
Type IIB
supergravity
D3-brane Dirac-Born-Infeld system, starting from the world-volume spike soliton
describing a
fundamental
string ending on the D3-brane \refs{\callan\gibbons{--}\howe}.
For this case, a useful ansatz for the full non-linear metric is not
known, but
we hope that our work below will help to motivate one.
The total action is given by $S=S_{bulk} + S_{kinetic} + S_{WZ}$,
where these terms are given in the Einstein frame by\refs{\schmid, \dt}
\eqn\actions{\eqalign{
&S_{bulk} = {1\over g_s^2}\int d^{10}x
\sqrt{g}\left[R-{1\over 2}\partial_\mu\phi\partial^\mu\phi -
{1\over 12}e^{-\phi}H^2
-\sum_n{1\over 2n!}e^{a_n\phi} {F^{(R)}_{[n]}}^2\right] \cr
&S_{kinetic} = -{1\over g_s}\int d^4 \xi  \sqrt{-\det(G_{ab} +
e^{-\phi/2} {\cal
F}_{ab})} \cr
&S_{WZ} = -{1\over 6g_s}\int D  - {1\over 4g_s}\int B^{R} \wedge {\cal F}
-
{1\over 8g_s}\int B^{NS} \wedge B^{R} - {1\over 16g_s}\int l {\cal F}
\wedge {\cal F} \ .\cr}}
Here, $B^{NS}$ is the NS-NS 2-form field, ${D,B^{R},l}$ are the RR
4, 2 and 0-form fields and $\phi$ is the dilaton. We have also defined
$G_{ab} = \partial_a X^{\mu}\partial_b X^{\nu} g_{\mu \nu}$ and
${\cal F}_{ab} =  F_{ab} -
\partial_a X^{\mu}\partial_b X^{\nu}B^{NS}_{\mu\nu}$,
where $F_{ab}$ is the field strength associated with the $U(1)$
connection
$A_{a}$
living on the brane. The field $H=dB^{(NS)}$ is the NS-NS field
strength,
$F_{[n]}$ are the field strengths of the
corresponding RR gauge potentials, and $a_1=2$, $a_3=1$, $a_5=0$.
We have also set $l_s=1$. Recall that  $F_[5]$ is a self--dual field strenght. This information cannot be inserted in a covariant action, and therefore we must have in mind that the complete solution for $F_{[5]}$ in terms of the field $D$ in our equations is $F_{[5]}=dD+ ^{*}dD$.

In the weak coupling limit $g_s \rightarrow 0$, the field equations to
zeroth
order in $g_s$
are satisfied by the world volume spike soliton
\refs{\callan\gibbons{--}\howe},
representing
a fixed number $N_F$ of fundamental strings ending on the D3-brane. in
the flat background
$g_{\mu\nu}=\eta_{\mu\nu}$ with all other the bulk fields equal to zero,
given
in static gauge
$\xi^a= X^a=x^a$, $a=0,1,2,3$ by
\eqn\parametr{
X^{9} = {\alpha^2 \over r},\qquad  A_{0} = {\alpha^2\over r} ,}
where $r^2 = x_1^2 +x_2^2 +x_3^2$ and $\alpha^2 = g_s N_F $.
Although it looks much like the parameter $\alpha_0$ associated
with the intersecting brane solutions of sections 2 and 3, the parameter
$\alpha$ appearing here is physically much different.  It does not
correspond
to a modulus in the field theory and in fact is quantized since the
number
$N_F$ of fundamental string charge must be an integer.

Our aim is to linearize the bulk field equations and compute the first
order
corrections in $g_s$.  The form \parametr\ solves the Born-Infeld
equations
only in the limit of small $\alpha$ \callan,
but this is achieved for $g_s \rightarrow 0$ with
$N_F$ fixed.  Now, small $\alpha$ will in fact mean that, for example,
the
extrinsic curvature of the embedded 3+1 surface will be large.  In
general,
we would not expect even the exact Born-Infeld description to be valid
in this
domain.  Luckily, for intersections of this form, it was shown in \LT\
that the
Born-Infeld description is in fact exact.

The non-zero components of the brane stress tensor are given by
\eqn\stress{\eqalign{
&T_{brane}^{00} = {1\over 2g_s} \left(1+{\alpha^4\over
r^4}\right)\delta^{(6)}, \qquad T_{brane}^{ii} = -{1\over 2g_s}
\delta^{(6)}\cr
&T_{brane}^{i9}=T_{brane}^{9i} = {\alpha^2\over 2g_s} {x^i\over
r^3}\delta^{(6)}, \qquad
T_{brane}^{99} = -{1\over 2g_s} {\alpha^4\over r^4} \delta^{(6)} \ ,
\cr}}
where the index $i$ ranges over $1,2,3$ and
$\delta^{(6)} = \delta(x^4)\delta(x^5) \cdots
\delta(x^8)\delta\left(x^9-{\alpha^2\over r}\right)$ .
The  final
expression for $h_{\mu\nu}$ can be given in terms of the following three
integrals:
\eqn\fq{\eqalign{
&f_{0}(x) = {1 \over 7 \omega_8}\int  {d^3 x'\over
\left[(x_1-x'_1)^2 +
\cdots
+(x_3-x'_3)^2 +
x_4^2 + \cdots x_8^2 + (x_9-{\alpha^2\over r'})^2\right]^{7/2}} , \cr
&f^i_{1}(x) = {1 \over 7 \omega_8}\int  {{x'}^{i}d^3 x'\over
{r'}^{3}\left[(x_1-x'_1)^2 +
\cdots
+(x_3-x'_3)^2 +
x_4^2 + \cdots x_8^2 + (x_9-{\alpha^2\over r'})^2\right]^{7/2}} , \cr
&f_{2}(x) = {1 \over 7 \omega_8}\int  {d^3 x'\over
{r'}^{4}\left[(x_1-x'_1)^2 +
\cdots
+(x_3-x'_3)^2 +
x_4^2 + \cdots x_8^2 + (x_9-{\alpha^2\over r'})^2\right]^{7/2}} ,
\cr}}
with $\omega_8$ being the area of the $8$--sphere.  The
solution for the linearized Einstein metric is
\eqn\hh{\eqalign{
&h_{00} = {g_s\over 2}(f_0+ {3\alpha^4\over 2} f_2), \qquad
h_{ii} = -{g_s\over 2}(f_0 -
{\alpha^4\over
2} f_2) \cr
&h_{AA} = {g_s\over 2}(f_0+ {\alpha^4\over 2} f_2),\qquad
h_{99} = {g_s\over 2}(f_0 - {3 \alpha^4\over 2} f_2),\qquad
h_{9i}=h_{i9} =
g_s\alpha^2 f^i_1 \ ,
\cr}}
where $A=4,\ldots 8$.

Varying the action with respect to the dilaton and keeping only terms
that are first order in $g_s$ yelds to
$\partial^\mu \partial_\mu \phi =  {g_s\alpha^4\over 2  r^4} \delta^{(6)}$
which has the solution
\eqn\dilaton{
\phi =-{1\over 2}g_s\alpha^4 f_2  \ .
}
For the NS 3-form field strength we have the linearized equation
$g_s^{-2}\partial_{\mu}H^{\mu\alpha\beta} = J_{(NS)}^{\alpha\beta}$
with nonzero current components
\eqn\currentsns{
J_{(NS)}^{0i}=-{\alpha^2 x^i \over g_s r^3}\delta^{(6)},\qquad
J_{(NS)}^{09}= {\alpha^4\over g_sr^4}\delta^{(6)}.}
In the ``Lorentz Gauge", these equations read simply,
$g_s^{-2}\partial^{\mu}\partial_{\mu} B_{(NS)}^{\alpha\beta} =
J_{(NS)}^{\alpha\beta}$
and have the solution
\eqn\solns{
B^{(NS)}_{0i}=  -\alpha^2 g_s f^i_1,\qquad
B^{(NS)}_{09}=  \alpha^4 g_s  f_2 \ ,}
with all other components vanishing.  These are exactly the bulk gauge
fields that would be excited by a fundamental string aligned in the
$x^9$
direction.

%%%%%%%%%%%%%%%%%%%%%%%%%%%%%%%%%%%%%%%%%%%%%%%%%%%
%%%%%%%%%%%%%%%%%%%%%%%%%%%%%%%%%%%%%%%%%%%%%%%%%%%

The first order equations for the RR fields are
\eqn\rreq{
 {1\over g_s^2} \nabla_\mu {F_{[n]}^{(R)}}^{\mu\alpha\cdots\gamma} =
{J^{(R)}}^{\alpha\cdots\gamma} \ ,
 }
where the only  non--zero currents are,
\eqn\currentrf{
%\eqalign{
J_{(R)}^{ij} = -{\alpha^2 x_k \epsilon^{ijk}\over
32 g_s r^3} 
\delta^{(6)},\quad
J^{0ijk} = {\epsilon^{ijk}\over 144 g_s}\delta^{(6)},\quad
J^{09jk} = -{\alpha^2 \epsilon^{ijk} x_i \over 144 g_s r^3}\delta^{(6)} \ ,
}
and the components obtained by permutations of their indices.
The current associated to the $0$--form $l$ vanishes.
Again we use the ``Lorentz Gauge'' to solve the equations, and  obtain
\eqn\solr{
%\eqalign{
B^{(R)}_{ij} =  {\alpha^2  g_s \over 32}\epsilon_{ijk}f^k_1 ,\qquad
 D_{0ijk} = {g_s\over 144}\epsilon_{ijk}f_0 , \qquad
D_{09jk} = -{g_s \alpha^2 \over 144} \epsilon_{ijk}f^i_1} .

%%%%%%%%%%%%%%%%%%%%%%%%%%%%%%%%%%%%%%%%%%%%%
%%%%%%%%%%%%%%%%%%%%%%%%%%%%%%%%%%%%%%%%%%%%%

Let us now explore the form of the integrals $f$ in \fq. Considcr $f_q$, with $q=0,2$. The symmetries of
all the expressions show that we can rotate the $x^i$ plane and the
$(x_4,\ldots,x_8)$ plane in such a way that any point in spacetime is
equivalent
to one such the only nonzero components are $x_3$, $x_8$ and $x_9$. In
that
situation we integrate over $\theta$ obtaining,
\eqn\fq{\eqalign{
&f_q(x_3,x_8,x_9)
= {4\pi\over 5 x_3 \omega_8}\int_0^{\infty}dr
r^{2(3-q)}\left[{1\over
(r^2[(r-x_3)^2+x_8^2]+[\alpha^2-rx_9]^2)^{5/2}}\right.
\cr
&\left.  \hskip 5cm - {1\over
(r^2[(r+x_3)^2+x_8^2]+[\alpha^2-rx_9]^2)^{5/2}}\right]  \ .
\cr}}
It is easy to see that these integrals will be convergent. For
$r\rightarrow \infty$ they go like $\int{dr \ r^{-2(q+2)}}$. Note that
when
$x_3\rightarrow 0$ both (infinite) terms in \fq{}  cancel each other.
The
only singularity occurs when $x$ is located over the source, that is,
when
\eqn\ond{
x_8=0,\qquad x_3 x_9 = \pm \alpha^2 .
}
This was expected, and means that our perturbative analysis is not valid
near the source. The figure shows a plot of $7 \omega_8 f_2$
(which corresponds to the dilaton) with fixed $x_8=0$ and $\alpha=1$.
The plot was made by evaluating the integral \fq{} numerically.

\vskip 1cm
\epsfxsize=10cm
\leavevmode
\epsfbox{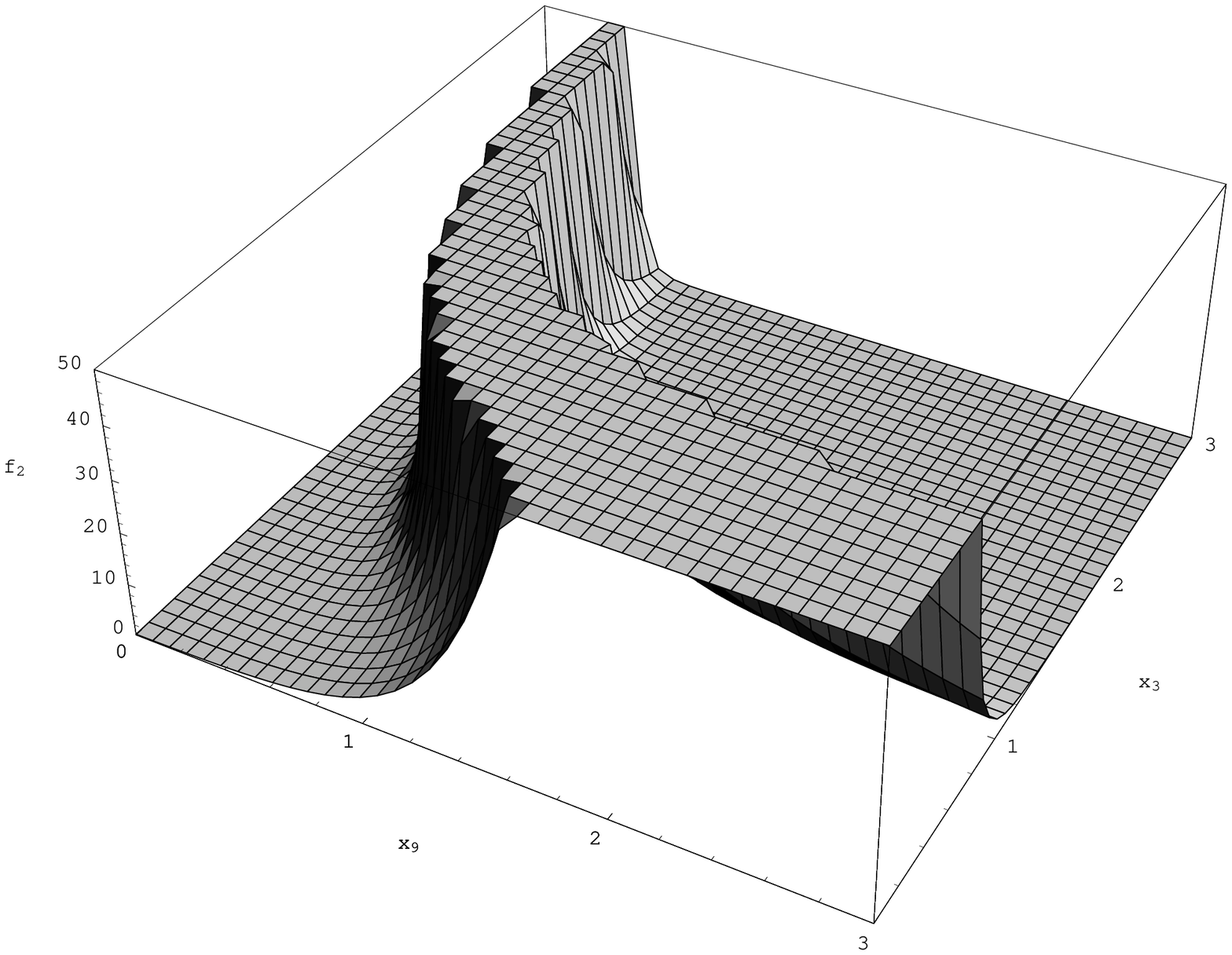}

The flat region is a ``numerical cutoff'' near the singularity at the
source; i.e., it is just the region where $\omega_8 f_2 \ge 50$.
Note how the singular region narrows, indicating a weaker singularity,
far from $x_3=0$.   Recall that large $x_3$ is far from the fundamental
string.
This behavior is therefore expected, since we know that
a pure D3-brane by itself is not a source for the dilaton.
The other functions $f_q$ and $k_q$ show similar behavior.

Here we have studied only the lowest order bulk fields in the limit of
small $\alpha$.  It would be interesting both to understand the first
order
fields produced by the exact BI-on solution \callan\ and
to study higher order contributions to the bulk fields.  For the
case where the string passes through the D3-brane (and does not end on
it),
\amdon\ would again predict that a fully localized intersecting brane
solution
does not exist.   The argument involves considering the S-dual system of
a D1-brane intersecting a D3-brane and identifying a set of moduli which
live on the 0+1 dimensional intersection manifold and which
are T-dual to the moduli that determine the delocalization of the
D$2\perp$D$2(0)$
intersection.  In this case, these moduli are not associate with
the parameter $\alpha$, but rather with the fact that the two halves of
the
string on opposite sides of the D3-brane can separate.
Note, however, that the case considered here
is somewhat different since we only have a string on a single side
of the D3-brane. In particular, we cannot consider this solution
as a limit of solutions in which the branes are separated in a
transverse
direction.
Therefore, it appears possible that the present case may have different
behavior.

\vskip 1cm

\listrefs
\end